\documentclass[english,conference]{IEEEtran}
\usepackage[T1]{fontenc}
\usepackage[latin9]{inputenc}
\usepackage{graphicx}

\makeatletter
\IEEEoverridecommandlockouts
\usepackage{url}
\usepackage[pdftex,pdftitle=A Hybrid DSP/Deep Learning Approach to Real-Time Full-Band Speech Enhancement,pdfauthor=Jean-Marc Valin]{hyperref}

\makeatother

\usepackage{babel}
\begin{document}


\title{A Hybrid DSP/Deep Learning Approach to Real-Time Full-Band Speech
Enhancement}

\author{\IEEEauthorblockN{Jean-Marc Valin}\IEEEauthorblockA{Mozilla Corporation\\Mountain View, CA, USA\\\href{mailto:jmvalin@jmvalin.ca}{jmvalin@jmvalin.ca}}}
\maketitle
\begin{abstract}
Despite noise suppression being a mature area in signal processing,
it remains highly dependent on fine tuning of estimator algorithms
and parameters. In this paper, we demonstrate a hybrid DSP/deep learning
approach to noise suppression. We focus strongly on keeping the complexity
as low as possible, while still achieving high-quality enhanced speech.
A deep recurrent neural network with four hidden layers is used to
estimate ideal critical band gains, while a more traditional pitch
filter attenuates noise between pitch harmonics. The approach achieves
significantly higher quality than a traditional minimum mean squared
error spectral estimator, while keeping the complexity low enough
for real-time operation at 48~kHz on a low-power CPU. 
\end{abstract}


\section{Introduction}

\begin{figure}
\centering{\includegraphics[width=0.95\columnwidth]{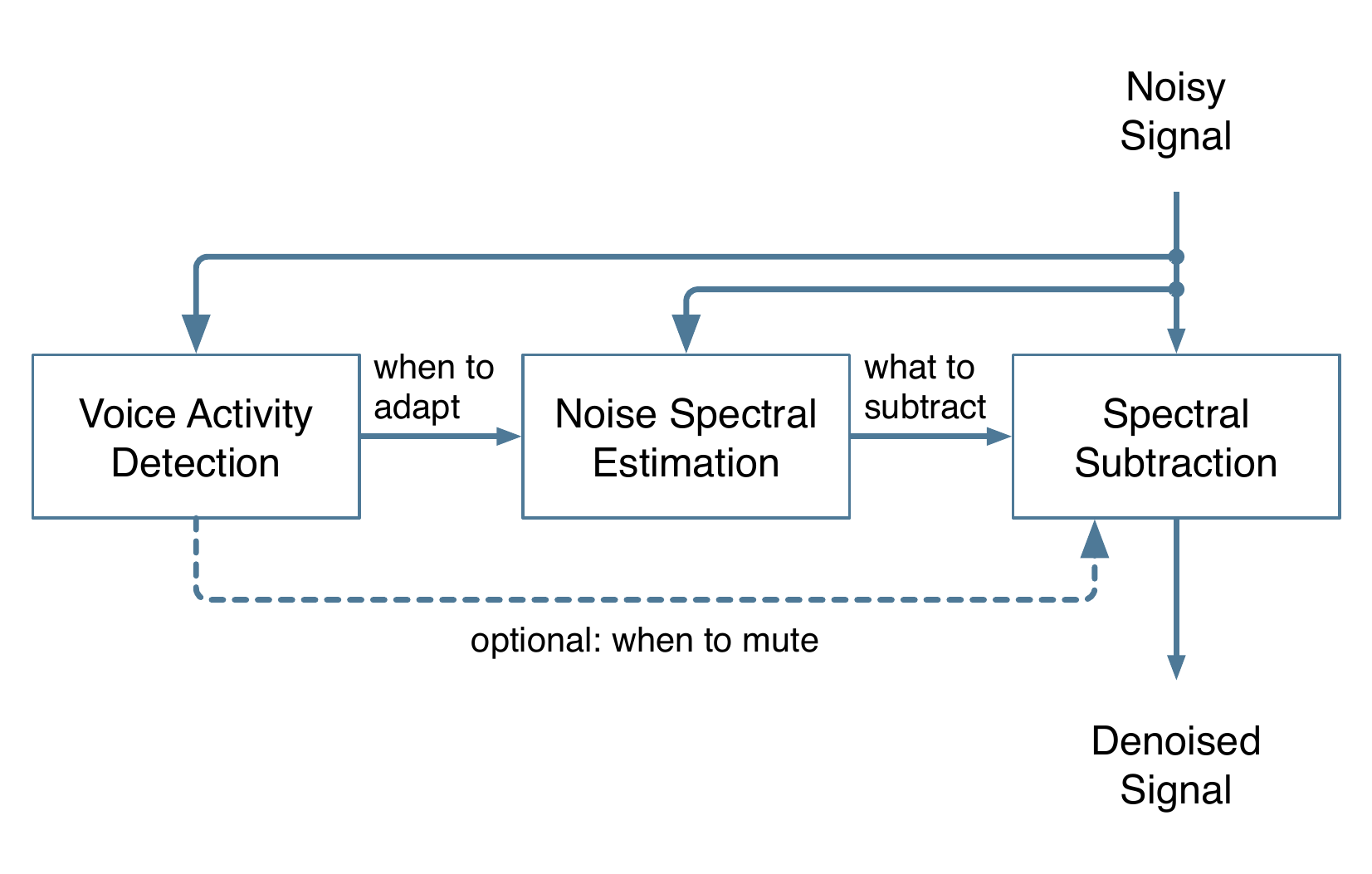}}

\caption{High-level structure of most noise suppression algorithms.\label{fig:High-level-structure}}
\end{figure}

Noise suppression has been a topic of interest since at least the
70s. Despite significant improvements in quality, the high-level structure
has remained mostly the same. Some form of spectral estimation technique
relies on a noise spectral estimator, itself driven by a voice activity
detector (VAD) or similar algorithm, as shown in Fig.~\ref{fig:High-level-structure}.
Each of the 3~components requires accurate estimators and are difficult
to tune. For example, the crude initial noise estimators and the spectral
estimators based on spectral subtraction~\cite{boll1979suppression}
have been replaced by more accurate noise estimators~\cite{hirsch1995noise,gerkmann2012unbiased}
and spectral amplitude estimators~\cite{ephraim1985speech}. Despite
the improvements, these estimators have remained difficult to design
and have required significant manual tuning effort. That is why recent
advances in deep learning techniques are appealing for noise suppression. 

Deep learning techniques are already being used for noise suppression~\cite{GoogleNS,liu2014experiments,xu2015regression,narayanan2013ideal,mirsamadi2016causal}.
Many of the proposed approaches target automatic speech recognition
(ASR) applications, where low latency is not required. Also, in many
cases, the large size of the neural network makes a real-time implementation
difficult without a GPU. In the proposed approach we focus on real-time
applications (e.g. video-conference) with low complexity. We also
focus on full-band (48~kHz) speech. To achieve these goals we choose
a hybrid approach (Sec.~\ref{sec:Signal-Model}), where we rely on
proven signal processing techniques and use deep learning (Sec.~\ref{sec:Deep-Learning-Architecture})
to replace the estimators that have traditionally been hard to correctly
tune. The approach contrasts with so-called \emph{end-to-end} systems
where most or all of the signal processing operations are replaced
by machine learning. These end-to-end system have clearly demonstrated
the capabilities of deep learning, but they often come at the cost
of significantly increased complexity.

We show that the proposed approach has an acceptable complexity (Sec.~\ref{sec:Complexity-Analysis})
and that it provides better quality than more conventional approaches
(Sec.~\ref{sec:Results}). We conclude in Sec.~\ref{sec:Conclusion}
with directions for further improvements to this approach.

\section{Signal Model}

\label{sec:Signal-Model}

We propose a hybrid approach to noise suppression. The goal is to
use deep learning for the aspects of noise suppression that require
careful tuning while using basic signal processing building blocks
for parts that do not. 

The main processing loop is based on 20~ms windows with 50\%~overlap
(10~ms offset). Both analysis and synthesis use the same Vorbis window~\cite{montgomery2004vorbis},
which satisfies the Princen-Bradley criterion~\cite{princen1986analysis}.
The window is defined as
\begin{equation}
w\left(n\right)=\sin\left[\frac{\pi}{2}\sin^{2}\left(\frac{\pi n}{N}\right)\right]\ ,\label{eq:vorbis-window}
\end{equation}
where $N$ is the window length. 

\begin{figure}
\centering{\includegraphics[width=1\columnwidth]{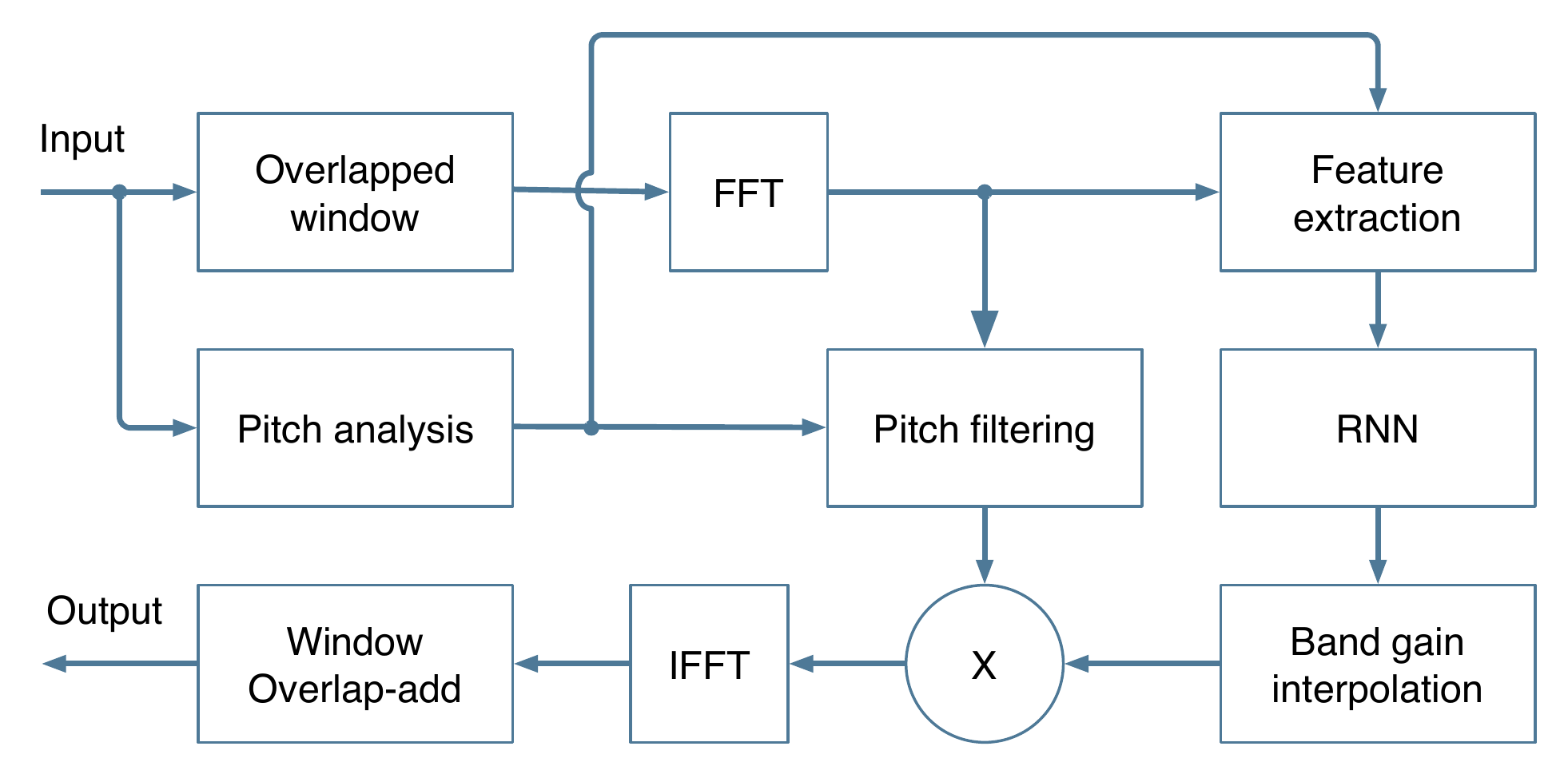}}

\caption{Block diagram.\label{fig:Block-diagram}}
\end{figure}

The signal-level block diagram for the system is shown in Fig.~\ref{fig:Block-diagram}.
The bulk of the suppression is performed on a low-resolution spectral
envelope using gains computed from a recurrent neural network (RNN).
Those gains are simply the square root of the ideal ratio mask~(IRM).
A finer suppression step attenuates the noise between pitch harmonics
using a pitch comb filter. 

\subsection{Band structure}

In the approach taken by \cite{GoogleNS}, a neural network is used
to directly estimate magnitudes of frequency bins and requires a total
of 6144~hidden units and close to 10~million weights to process
8~kHz speech. Scaling to 48~kHz speech using 20-ms frames would
require a network with 400~outputs (0~to 20~kHz), which clearly
results in a higher complexity than we can afford.

One way to avoid the problem is to assume that the spectral envelopes
of the speech and noise are sufficiently flat to use a coarser resolution
than frequency bins. Also, rather than directly estimating spectral
magnitudes, we instead estimate ideal critical band gains, which have
the significant advantage of being bounded between 0 and~1. We choose
to divide the spectrum into the same approximation of the Bark scale~\cite{moore2012introduction}
as the Opus codec~\cite{Valin2013} uses. That is, the bands follow
the Bark scale at high frequencies, but are always at least 4~bins
at low frequencies. Rather than rectangular bands, we use triangular
bands, with the peak response being at the boundary between bands.
This results in a total of 22~bands. Our network therefore requires
only 22~output values in the $\left[0,1\right]$ range. 

Let $w_{b}(k)$ be the amplitude of band $b$ at frequency $k$, we
have $\sum_{b}w_{b}\left(k\right)=1$. For a transformed signal $X\left(k\right)$,
the energy in a band is given by
\begin{equation}
E\left(b\right)=\sum_{k}w_{b}\left(k\right)\left|X\left(k\right)\right|^{2}\ .\label{eq:band-energy}
\end{equation}
The per-band gain is defined as $g_{b}$
\begin{equation}
g_{b}=\sqrt{\frac{E_{s}\left(b\right)}{E_{x}\left(b\right)}}\ ,\label{eq:IRM}
\end{equation}
where $E_{s}\left(b\right)$ is the energy of the clean (ground truth)
speech and $E_{x}\left(b\right)$ is the energy of the input (noisy)
speech. Considering an ideal band gain $\hat{g}_{b}$, the following
interpolated gain is applied to each frequency bin $k$:
\begin{equation}
r\left(k\right)=\sum_{b}w_{b}\left(k\right)\hat{g}_{b}\ .\label{eq:band-interpolation}
\end{equation}

\subsection{Pitch filtering}

The main disadvantage of using Bark-derived bands to compute the gain
is that we cannot model finer details in the spectrum. In practice,
this prevents noise suppression between pitch harmonics. As an alternative,
we can use a comb filter at the pitch period to cancel the inter-harmonic
noise in a similar way that speech codec post-filters operate~\cite{chen1995adaptive}.
Since the periodicity of speech signal depends heavily on frequency
(especially for 48~kHz sampling rate), the pitch filter operates
in the frequency domain based on a per-band filtering coefficient
$\alpha_{b}$. Let $P(k)$ be the windowed DFT of the pitch-delayed
signal $x\left(n-T\right)$, the filtering is performed by computing
$X\left(k\right)+\alpha_{b}P\left(k\right)$ and then renormalizing
the resulting signal to have the same energy in each band as the original
signal $X\left(k\right)$. 

The pitch correlation for band $b$ is defined as
\begin{equation}
p_{b}=\frac{\sum_{k}w_{b}\left(k\right)\Re\left[X\left(k\right)P^{*}\left(k\right)\right]}{\sqrt{\sum_{k}w_{b}\left(k\right)\left|X\left(k\right)\right|^{2}\cdot\sum_{k}w_{b}\left(k\right)\left|P\left(k\right)\right|^{2}}}\ ,\label{eq:pitch-corr}
\end{equation}
where $\Re\left[\cdot\right]$ denotes the real part of a complex
value and $\cdot^{*}$ denotes the complex conjugate. Note that for
a single band, (\ref{eq:pitch-corr}) would be equivalent to the time-domain
pitch correlation. 

Deriving the optimal values for the filtering coefficient $\alpha_{b}$
is hard and the values that minimize mean squared error are not perceptually
optimal. Instead, we use a heuristic based on the following constraints
and observations. Since noise causes a decrease in the pitch correlation,
we do not expect $p_{b}$ to be greater than $g_{b}$ \emph{on average},
so for any band that has $p_{b}\geq g_{b}$, we use $\alpha_{b}$=1
. When there is no noise, we do not want to distort the signal, so
when $g_{b}=1$, we use $\alpha_{b}=0$. Similarly, when $p_{b}=0$,
we have no pitch to enhance, so $\alpha_{b}=0$. Using the following
expression for the filtering coefficient respects all these constraints
with smooth behavior between them:
\begin{equation}
\alpha_{b}=\min\left(\sqrt{\frac{p_{b}^{2}\left(1-g_{b}^{2}\right)}{\left(1-p_{b}^{2}\right)g_{b}^{2}}},\ 1\right)\ .\label{eq:pitch-filter-coeff}
\end{equation}

Even though we use an FIR pitch filter here, it is also possible to
compute $P(k)$ based on an IIR pitch filter of the form $H(z)=1/\left(1-\beta z^{-T}\right)$,
resulting in more attenuation between harmonics at the cost of slightly
increased distortion. 

\subsection{Feature extraction}

It only makes sense for the input of the network to include the log
spectrum of the noisy signal based on the same bands used for the
output. To improve the conditioning of the training data, we apply
a DCT on the log spectrum, which results in 22~Bark-frequency cepstral
coefficients (BFCC). In addition to these, we also include the temporal
derivative and the second temporal derivative of the first 6~BFCCs.
Since we already need to compute the pitch in~(\ref{eq:pitch-corr}),
we compute the DCT of the pitch correlation across frequency bands
and include the first 6~coefficients in our set of features. At last,
we include the pitch period as well as a spectral non-stationarity
metric that can help in speech detection. In total we use 42~input
features.

Unlike the features typically used in speech recognition, these features
do not use cepstral mean normalization and do include the first cepstral
coefficient. The choice is deliberate given that we have to track
the absolute level of the noise, but it does make the features sensitive
to the absolute amplitude of the signal and to the channel frequency
response. This is addressed in Sec.~\ref{subsec:training-data}.

\section{Deep Learning Architecture}

\label{sec:Deep-Learning-Architecture}

\begin{figure}
\centering{\includegraphics[width=0.95\columnwidth]{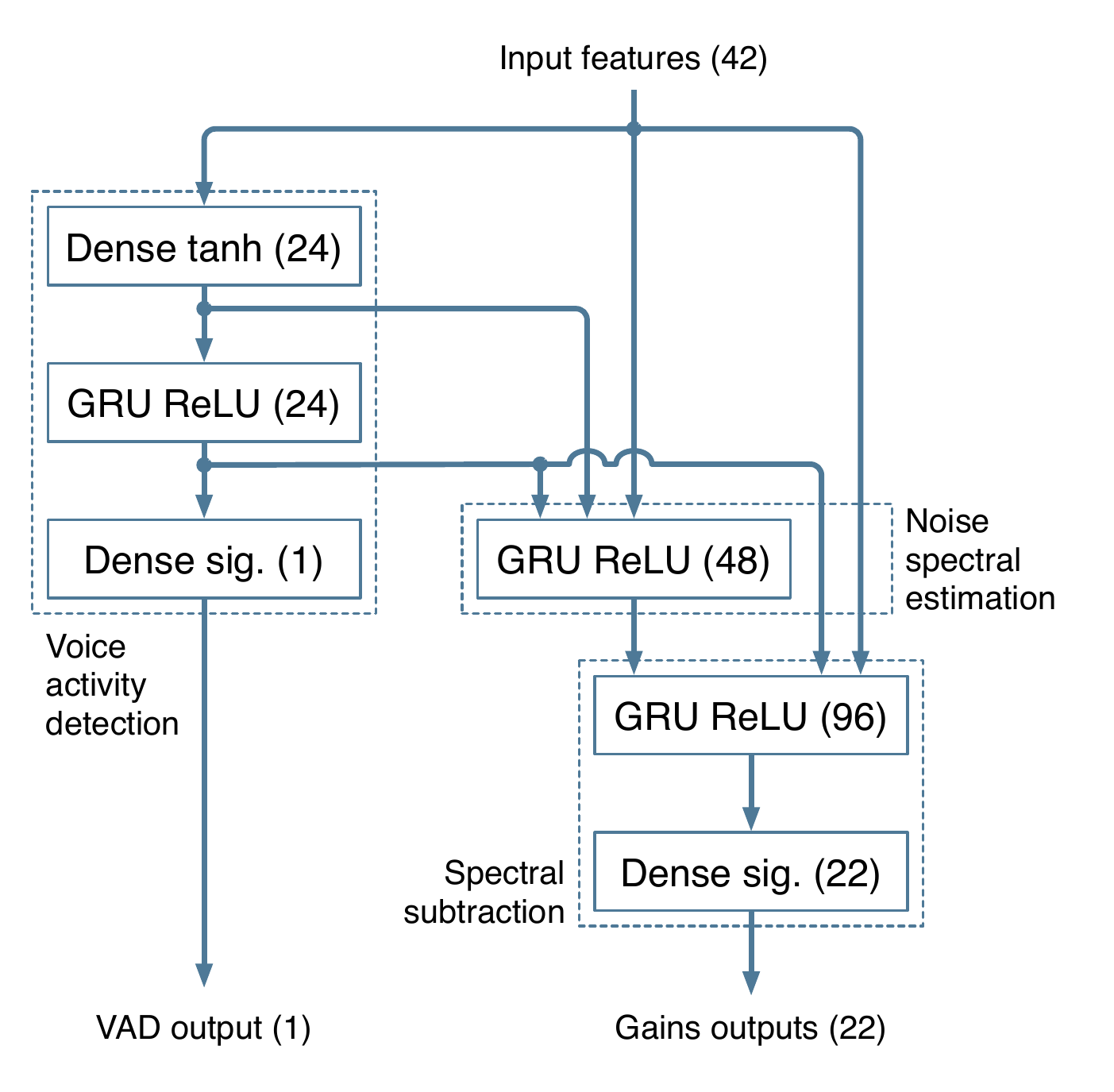}}

\caption{Architecture of the neural network, showing the feed-forward, fully
connected (dense) layers and the recurrent layers, along with the
activation function and the number of units for each layer.\label{fig:Architecture-of-the-network} }
\end{figure}

The neural network closely follows the traditional structure of noise
suppression algorithms, as shown in Fig.~\ref{fig:Architecture-of-the-network}.
The design is based on the assumption that the three recurrent layers
are each responsible for one of the basic components from Fig.~\ref{fig:High-level-structure}.
Of course, in practice the neural network is free to deviate from
this assumption (and likely does to some extent). It includes a total
of 215~units, 4~hidden layers, with the largest layer having 96~units.
We find that increasing the number of units does not significantly
improve the quality of the noise suppression. However, the loss function
and the way we construct the training data have a large impact on
the final quality. We find that gated recurrent unit (GRU)~\cite{cho2014properties}
slightly outperforms LSTM on this task, while also being simpler. 

Despite the fact that it is not strictly necessary, the network includes
a VAD output. The extra complexity cost is very small (24~additional
weights) and it improves training by ensuring that the corresponding
GRU indeed learns to discriminate speech from noise.

\subsection{Training data}

\label{subsec:training-data}

Since the ground truth for the gains requires both the noisy speech
and the clean speech, the training data has to be constructed artificially
by adding noise to clean speech data. For speech data, we use the
McGill TSP speech database\footnote{\url{http://www-mmsp.ece.mcgill.ca/Documents/Data/}}
(French and English) and the NTT Multi-Lingual Speech Database for
Telephonometry\footnote{The 44.1~kHz audio CD tracks are used rather than the 16~kHz data
files.} (21~languages). Various sources of noise are used, including computer
fans, office, crowd, airplane, car, train, construction. The noise
is mixed at different levels to produce a wide range of signal-to-noise
ratios, including clean speech and noise-only segments.

Since we do not use cepstral mean normalization, we use data augmentation
to make the network robust to variations in frequency responses. This
is achieved by filtering the noise and speech signal independently
for each training example using a second order filter of the form
\begin{equation}
H(z)=\frac{1+r_{1}z^{-1}+r_{2}z^{-2}}{1+r_{3}z^{-1}+r_{4}z^{-2}}\ ,\label{eq:random-filter}
\end{equation}
where each of $r_{1}\ldots r_{4}$ are random values uniformly distributed
in the $\left[-\frac{3}{8},\frac{3}{8}\right]$ range. Robustness
to the signal amplitude is achieved by varying the final level of
the mixed signal.

We have a total of 6~hours of speech and 4~hours of noise data,
which we use to produce 140~hours of noisy speech by using various
combinations of gains and filters and by resampling the data to frequencies
between 40~kHz and 54~kHz.

\subsection{Optimization process}

The loss function used for training determines how the network weighs
excessive attenuation versus insufficient attenuation when it cannot
exactly determine the correct gains. Although it is common to use
the binary cross-entropy function when optimizing for values in the
$\left[0,1\right]$ range, this does not produce good results for
the gains because it does not match their perceptual effect. For a
gain estimate $\hat{g}_{b}$ and the corresponding ground truth $g_{b}$,
we instead train with the loss function
\begin{equation}
L\left(g_{b},\hat{g}_{b}\right)=\left(g_{b}^{\gamma}-\hat{g}_{b}^{\gamma}\right)^{2}\ ,\label{eq:loss-gamma}
\end{equation}
where the exponent $\gamma$ is a perceptual parameter that controls
how aggressively to suppress noise. Since $\lim_{\gamma\rightarrow0}\frac{x^{\gamma}-1}{\gamma}=\log\left(x\right)$,
$\lim_{\gamma\rightarrow0}L\left(g_{b},\hat{g}_{b}\right)$ minimizes
the mean-squared error on the log-energy, which would make the suppression
too aggressive given the lack of a floor on the gain. In practice,
the value $\gamma=1/2$ provides a good trade-off and is equivalent
to minimizing the mean squared error on the energy raised to the power
$1/4$. Sometimes, there may be no noise and no speech in a particular
band. This is common either when the input is silent or at high frequency
when the signal is low-pass filtered. When that happens, the ground
truth gain is explicitly marked as undefined and the loss function
for that gain is ignored to avoid hurting the training process. 

For the VAD output of the network, we use the standard cross-entropy
loss function. Training is performed using the Keras\footnote{\url{https://keras.io/}}
library with the Tensorflow\footnote{\url{https://www.tensorflow.org/}}
backend. 

\subsection{Gain smoothing}

\begin{figure*}
\centering{\includegraphics[width=0.8\textwidth]{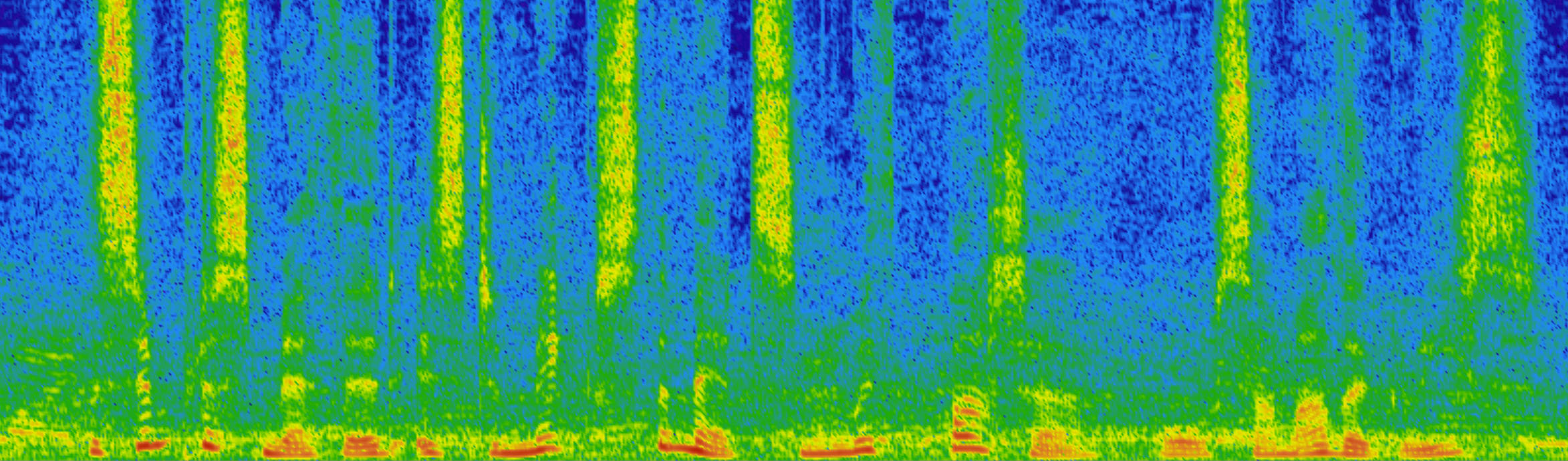}

\vspace{0.5em}

\includegraphics[width=0.8\textwidth]{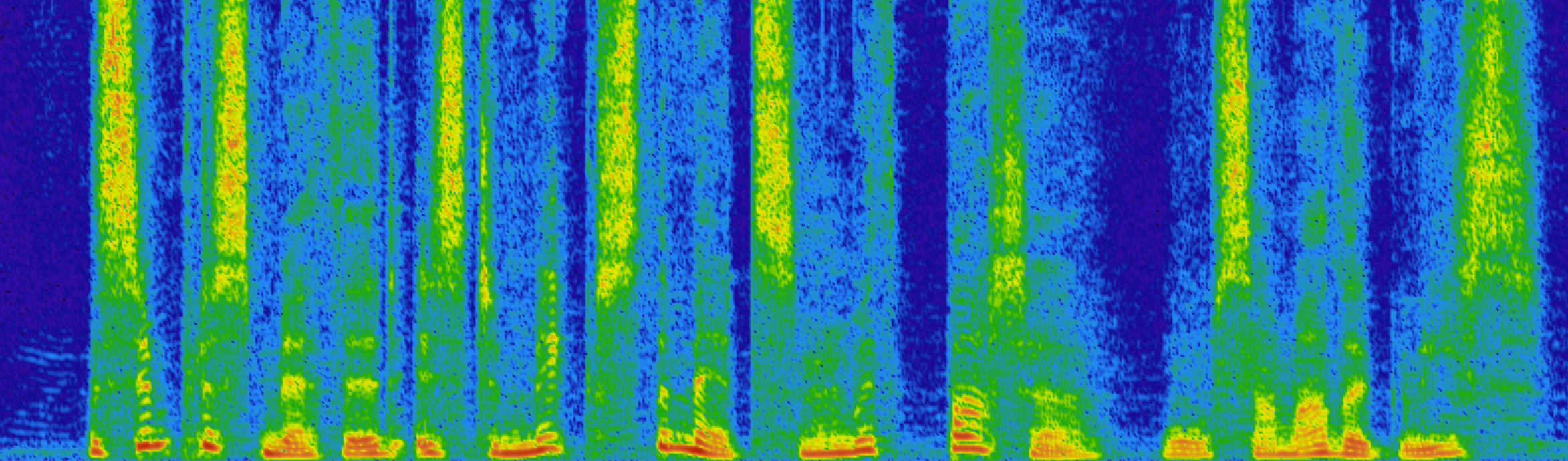}

\vspace{0.5em}

\includegraphics[width=0.8\textwidth]{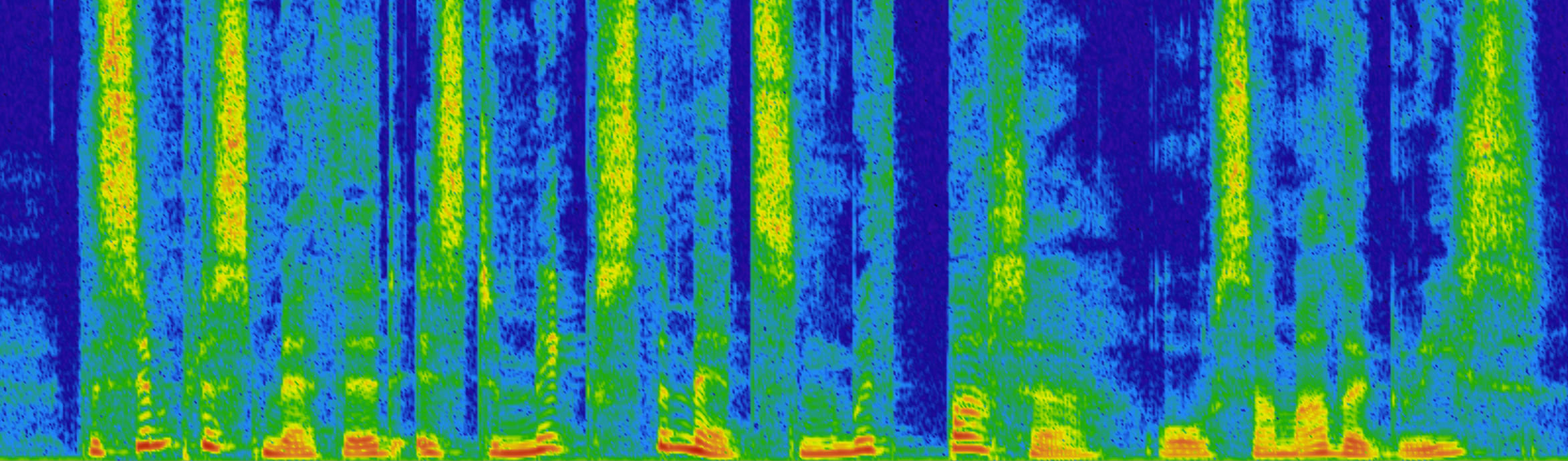}}

\caption{Example of noise suppression for babble noise at 15~dB SNR. Spectrogram
of the noisy (top), denoised (middle), and clean (bottom) speech.
For the sake of clarity, only the 0-12~kHz band is shown. \label{fig:Example-of-noise-suppression}}
\end{figure*}

When using the gains $\hat{g}_{b}$ to suppress noise, the output
signal can sometimes sound overly \emph{dry}, lacking the minimum
expected level of reverberation. The problem is easily remedied by
limiting the decay of $\hat{g}_{b}$ across frames. The smoothed gains
$\tilde{g}_{b}$ are obtained as
\begin{equation}
\tilde{g}_{b}=\max\left(\lambda\tilde{g}_{b}^{(prev)},\ \hat{g}_{b}\right)\ ,\label{eq:gain-filtering}
\end{equation}
where $\tilde{g}_{b}^{(prev)}$ is the filtered gain of the previous
frame and the decay factor $\lambda=0.6$ is equivalent to a reverberation
time of 135~ms. 

\section{Complexity Analysis}

\label{sec:Complexity-Analysis}

To make it easy to deploy noise suppression algorithms, it is desirable
to keep both the size and the complexity low. The size of the executable
is dominated by the 87,503~weights needed to represent the 215~units
in the neural networks. To keep the size as small as possible, the
weights can be quantized to 8~bits with no loss of performance. This
makes it possible to fit all weights in the L2 cache of a CPU. 

Since each weight is used exactly once per frame in a multiply-add
operation, the neural network requires 175,000~floating-point operations
(we count a multiply-add as two operations) per frame, so 17.5~Mflops
for real-time use. The IFFT and the two FFTs per frame require around
7.5~Mflops and the pitch search (which operates at 12~kHz) requires
around 10~Mflops. The total complexity of the algorithm is around
40~Mflops, which is comparable to that of a full-band speech coder. 

A non-vectorized C implementation of the algorithm requires around
1.3\% of a single x86 core (Haswell i7-4800MQ) to perform 48~kHz
noise suppression of a single channel. The real-time complexity of
the same floating-point code on a 1.2~GHz ARM Cortex-A53 core (Raspberry
Pi~3) is 14\%. 

As a comparison, the 16~kHz speech enhancement approach in~\cite{mirsamadi2016causal}
uses 3~hidden layers, each with 2048~units. This requires 12.5~million
weights and results in a complexity of 1600~Mflops. Even if quantized
to 8~bits, the weights do not fit the cache of most CPUs, requiring
around 800~MB/s of memory bandwidth for real-time operation.

\section{Results}

\label{sec:Results}

\begin{figure*}
\includegraphics[width=0.333\textwidth]{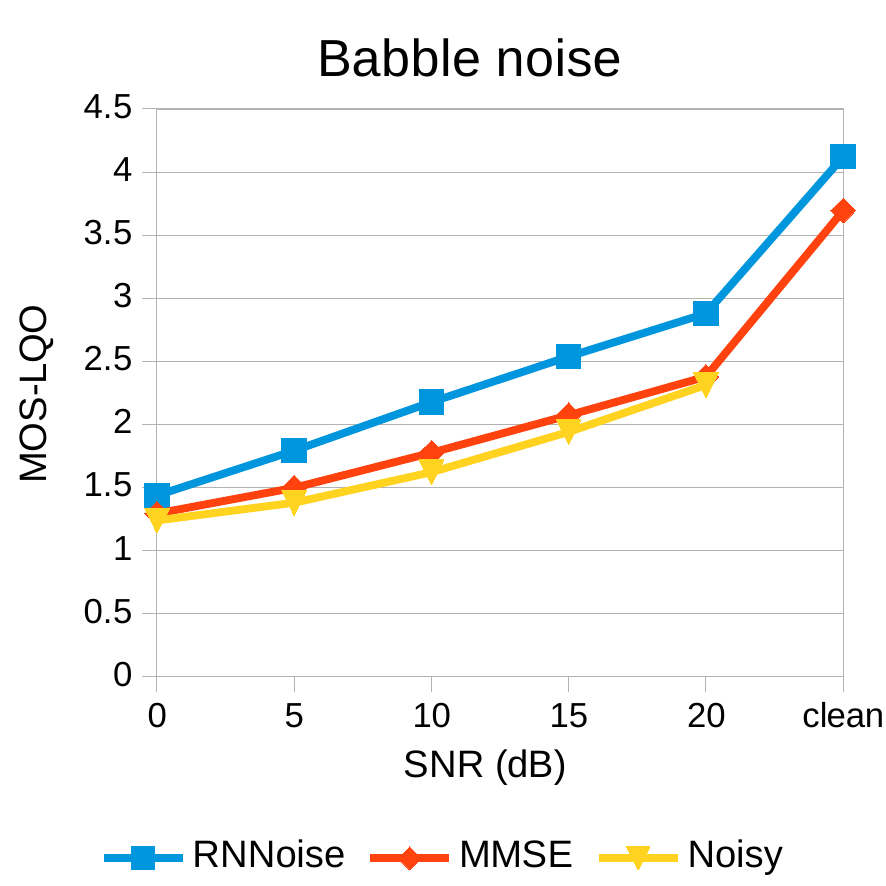}\includegraphics[width=0.333\textwidth]{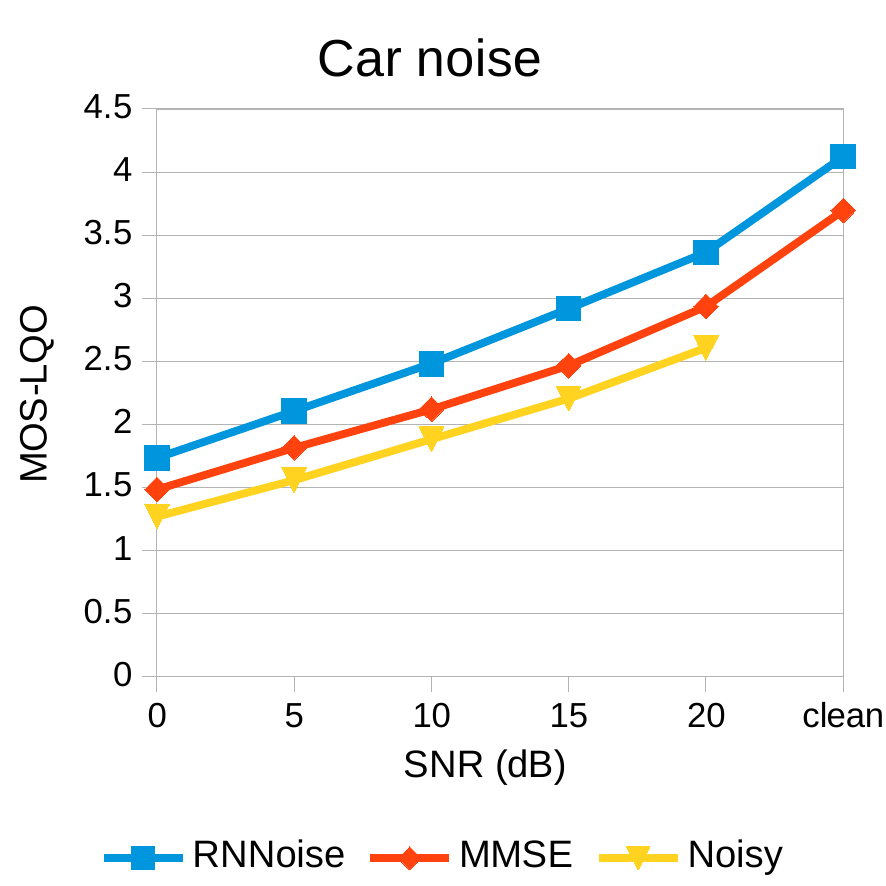}\includegraphics[width=0.333\textwidth]{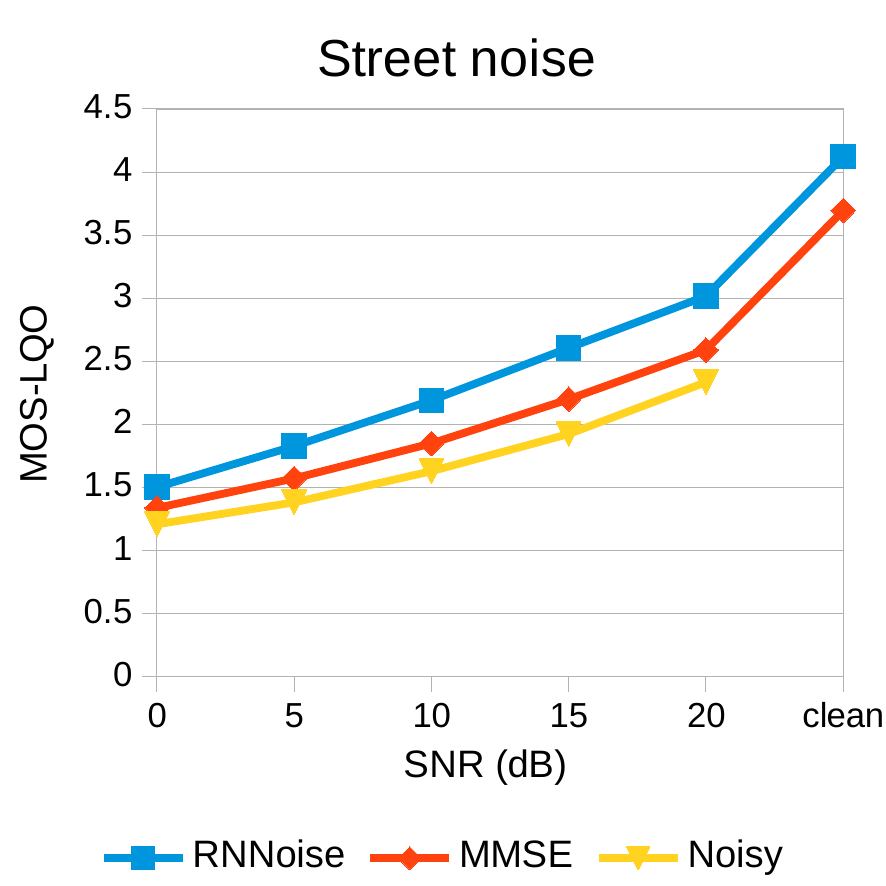}\caption{PESQ MOS-LQO quality evaluation for babble, car, and street noise.\label{fig:PESQ-MOS_LQO-quality}}
\end{figure*}

We test the quality of the noise suppression using speech and noise
data not used in the training set. We compare it to the MMSE-based
noise suppressor in the SpeexDSP library\footnote{\url{https://www.speex.org/downloads/}}.
Although the noise suppression operates at 48~kHz, the output has
to be resampled to 16~kHz due to the limitations of wideband PESQ~\cite{P.862}.
The objective results in Fig.~\ref{fig:PESQ-MOS_LQO-quality} show
a significant improvement in quality from the use of deep learning,
especially for non-stationary noise types. The improvement is confirmed
by casual listening of the samples. Fig.~\ref{fig:Example-of-noise-suppression}
shows the effect of the noise suppression on an example.

An interactive demonstration of the proposed system is available at
\mbox{\url{https://people.xiph.org/~jm/demo/rnnoise/}}, including
a real-time Javascript implementation. The software implementing the
proposed system can be obtained under a BSD license at \mbox{\url{https://github.com/xiph/rnnoise/}}
and the results were produced using commit hash \texttt{91ef401}. 

\section{Conclusion}

\label{sec:Conclusion}

This paper demonstrates a noise suppression approach that combines
DSP-based techniques with deep learning. By using deep learning only
for the aspects of noise suppression that are hard to tune, the problem
is simplified to computing only 22~ideal critical band gains, which
can be done efficiently using few units. The coarse resolution of
the bands is then addressed by using a simple pitch filter. The resulting
low complexity makes the approach suitable for use in mobile or embedded
devices and the latency is low enough for use in video-conferencing
systems. We also demonstrate that the quality is significantly higher
than that of a pure signal processing-based approach.

We believe that the technique can be easily extended to residual echo
suppression, for example by adding to the input features the cepstrum
of the far end signal or the filtered far-end signal. Similarly, it
should be applicable to microphone array post-filtering by augmenting
the input features with leakage estimates like in~\cite{valin2004microphone}.

\bibliographystyle{IEEEbib}
\bibliography{rnnoise}

\begin{thebibliography}{10}

\bibitem{boll1979suppression}
S.~Boll,
\newblock ``Suppression of acoustic noise in speech using spectral
  subtraction,''
\newblock {\em IEEE Transactions on acoustics, speech, and signal processing},
  vol. 27, no. 2, pp. 113--120, 1979.

\bibitem{hirsch1995noise}
H.-G. Hirsch and C.~Ehrlicher,
\newblock ``Noise estimation techniques for robust speech recognition,''
\newblock in {\em Proc. ICASSP}, 1995, vol.~1, pp. 153--156.

\bibitem{gerkmann2012unbiased}
T.~Gerkmann and R.C. Hendriks,
\newblock ``Unbiased {MMSE}-based noise power estimation with low complexity
  and low tracking delay,''
\newblock {\em IEEE Transactions on Audio, Speech, and Language Processing},
  vol. 20, no. 4, pp. 1383--1393, 2012.

\bibitem{ephraim1985speech}
Y.~Ephraim and D.~Malah,
\newblock ``Speech enhancement using a minimum mean-square error log-spectral
  amplitude estimator,''
\newblock {\em IEEE Transactions on Acoustics, Speech, and Signal Processing},
  vol. 33, no. 2, pp. 443--445, 1985.

\bibitem{GoogleNS}
A.~Maas, Q.V. Le, T.M. O'Neil, O.~Vinyals, P.~Nguyen, and A.Y. Ng,
\newblock ``Recurrent neural networks for noise reduction in robust {ASR},''
\newblock in {\em Proc. INTERSPEECH}, 2012.

\bibitem{liu2014experiments}
D.~Liu, P.~Smaragdis, and M.~Kim,
\newblock ``Experiments on deep learning for speech denoising,''
\newblock in {\em Proc. Fifteenth Annual Conference of the International Speech
  Communication Association}, 2014.

\bibitem{xu2015regression}
Y.~Xu, J.~Du, L.-R. Dai, and C.-H. Lee,
\newblock ``A regression approach to speech enhancement based on deep neural
  networks,''
\newblock {\em IEEE Transactions on Audio, Speech and Language Processing},
  vol. 23, no. 1, pp. 7--19, 2015.

\bibitem{narayanan2013ideal}
A.~Narayanan and D.~Wang,
\newblock ``Ideal ratio mask estimation using deep neural networks for robust
  speech recognition,''
\newblock in {\em Proc. ICASSP}, 2013, pp. 7092--7096.

\bibitem{mirsamadi2016causal}
S.~Mirsamadi and I.~Tashev,
\newblock ``Causal speech enhancement combining data-driven learning and
  suppression rule estimation.,''
\newblock in {\em Proc. INTERSPEECH}, 2016, pp. 2870--2874.

\bibitem{montgomery2004vorbis}
C.~Montgomery,
\newblock ``Vorbis {I} specification,'' 2004.

\bibitem{princen1986analysis}
J.~Princen and A.~Bradley,
\newblock ``Analysis/synthesis filter bank design based on time domain aliasing
  cancellation,''
\newblock {\em IEEE Tran. on Acoustics, Speech, and Signal Processing}, vol.
  34, no. 5, pp. 1153--1161, 1986.

\bibitem{moore2012introduction}
B.C.J. Moore,
\newblock {\em An introduction to the psychology of hearing},
\newblock Brill, 2012.

\bibitem{Valin2013}
J.-M. Valin, G.~Maxwell, T.~B. Terriberry, and K.~Vos,
\newblock ``High-quality, low-delay music coding in the {Opus} codec,''
\newblock in {\em Proc. 135th AES Convention}, 2013.

\bibitem{chen1995adaptive}
Juin-Hwey Chen and Allen Gersho,
\newblock ``Adaptive postfiltering for quality enhancement of coded speech,''
\newblock {\em IEEE Transactions on Speech and Audio Processing}, vol. 3, no.
  1, pp. 59--71, 1995.

\bibitem{cho2014properties}
K.~Cho, B.~Van~Merri{\"e}nboer, D.~Bahdanau, and Y.~Bengio,
\newblock ``On the properties of neural machine translation: Encoder-decoder
  approaches,''
\newblock in {\em Proc. Eighth Workshop on Syntax, Semantics and Structure in
  Statistical Translation (SSST-8)}, 2014.

\bibitem{P.862}
ITU-T,
\newblock {\em Perceptual evaluation of speech quality ({PESQ}): An objective
  method for end-to-end speech quality assessment of narrow-band telephone
  networks and speech codecs}, 2001.

\bibitem{valin2004microphone}
J.-M. Valin, J.~Rouat, and F.~Michaud,
\newblock ``Microphone array post-filter for separation of simultaneous
  non-stationary sources,''
\newblock in {\em Proc. ICASSP}, 2004.

\end{thebibliography}

\end{document}